\documentclass[aps,onecolumn]{revtex4-2}
\usepackage{graphicx}
\usepackage{epsfig}
\usepackage{epstopdf}
\usepackage{amsfonts}
\usepackage{amssymb}
\usepackage{amsbsy}
\usepackage{amsmath}
\usepackage{mathrsfs}
\usepackage{latexsym}
\usepackage{natbib}
\usepackage{bm}
\usepackage{color}
\usepackage[a4paper, total={6.3in, 9.9in}]{geometry}

\usepackage{xifthen}
\newcommand{\dalembert}[1][]{\ifthenelse{\isempty{#1}}{\Box}{#1\Box}}

\usepackage{ragged2e}
\usepackage{braket}
\usepackage{slashed}
\usepackage{pgfplots}
\numberwithin{equation}{section}

\usepackage{tikz}
\usetikzlibrary{shapes,arrows,shadows}
\begin{document}

\title{Neutrino oscillation in the presence of background classical sources}

\author{Susobhan Mandal}
\email{sm12ms085@gmail.com}

\affiliation{ Department of Physics, 
Indian Institute of Technology Bombay,
Mumbai - 400076, India }

\date{\today}

\begin{abstract}
\begin{center}
\underline{\textbf{Abstract}}
\end{center}
The presence of background classical sources affects a quantum field theory significantly in different ways. Neutrino oscillation is a phenomenon that confirms that neutrinos are massive fermions in nature, a celebrated result in modern physics. Neutrino oscillation plays an important role in many astrophysical observations. However, the interactions between the background classical sources with neutrinos are not often considered. In the present article, we show the effect of some classical sources, namely matter currents, electromagnetic waves, torsion, and gravitational waves on neutrino oscillation. It is shown explicitly that the above sources can change the helicity state of neutrinos during neutrino oscillation.%

\end{abstract}

\maketitle

\section{Introduction}

Neutrino oscillation is an interesting phenomenon in modern physics that is often observed in particle physics, nuclear physics \cite{bilenky1978lepton, gonzalez2001new, barger2000cpt, nunokawa2008cp}, and cosmology \cite{dolgov2002cosmological, kaplan2004neutrino, mandal2021neutrino, lesgourgues2006massive, akhmedov1998baryogenesis, canetti2013dark}. Neutrino oscillation is first suggested by Pontecorvo in order to confirm the massive nature of neutrinos \cite{pontecorvo1957mesonium}. After many years of studies and efforts, it has brought to the scientific community independent pieces of evidence of neutrino oscillations that have been obtained in a series of experiments \cite{fukuda1998evidence, fukuda1999measurement, fukuda2000tau, surdo2002atmospheric, giacomelli2004macro, ahmad2002direct, altmann2005complete, hosaka2006solar, abe2008precision, adamson2013measurement, ahn2006measurement}. The phenomenon of neutrino mixing and the existence of non-zero neutrino masses have captured much attention because it opens up fascinating possibilities in physics beyond the Standard Model.

The nature of transition probability in neutrino oscillation depends on the medium through which neutrinos are propagating. Hence, the interactions between medium and neutrinos affect the nature of neutrino oscillation, which has been studied in the literature \cite{nieves2018neutrino, duan2013flavor, abbar2018fast, wolfenstein1979effects, grigoriev2019neutrino}. Apart from the flavor oscillation, we also expect a change in helicity during the neutrino oscillation in a non-trivial medium, unlike in the vacuum neutrino oscillation. This has been discussed in the literature \cite{fabbricatore2016neutrino, dobrynina2016helicity, kartavtsev2015neutrino} considering different interactions. The foremost theme of the present article is to show the nature of neutrino oscillation in the presence of some background classical sources which include a conserved matter current. In order to show that, we used some results related to the Dirac-Volkov equations \cite{wolkow1935klasse}. Further, the change in the helicity state of neutrinos during the neutrino flavor oscillation is also shown explicitly in this case. Using a similar mathematical approach, neutrino oscillation in the presence of a background electromagnetic plane wave field is also shown, where neutrinos are coupled to the gauge field through their magnetic moments. This study would help us in understanding neutrino oscillations in the presence of astrophysical particles and other sources. 

General relativity (GR) is a successful theory of classical gravity that agrees with a vast number of observations. GR is also successful in predicting many astrophysical observations. Gravitational wave (GW) is one of the predictions of GR which is confirmed recently \cite{abbott2016gw151226, shandera2018gravitational, abbott2016observation, abbott2017gw170817}. The detection of GW is important as it is expected to open up new possibilities for both astrophysical and cosmological observations since the information carried by GW is different from its electromagnetic counterpart. This opens us to a new way of understanding black holes, their formation, the merging of binary systems consisting of compact astrophysical objects, the search for exotic compact astrophysical objects, and others. Since GW couples to matter sources through their stress-energy tensor, we expect it affects the nature of neutrino oscillation with a change in the helicity states of neutrinos. However, the change is quantitatively small as the strength of the coupling constant is very small. This is explicitly shown in the present article. Therefore, in principle, neutrino oscillation can also be used in detecting GW.

GR is described by the Einstein-Hilbert action, which yields the field equations known as Einstein's equations if varied with respect to the metric. However, if the metric and the non-Riemannian part of the connection, namely the torsion are considered as a priori independent variables assuming metric compatibility, then two field equations are obtained using the variational principle. The first one is nothing but the Einstein equation although the Einstein tensor, in this case, is not necessarily symmetric, while the other field equation relates the torsion tensor to the spin tensor. The existence of torsion promises to solve singularity and other problems in GR and cosmology, which are discussed in the literature \cite{kopczynski1972non, shapiro2002physical, kuchowicz1978friedmann, poplawski2010cosmology}. The existence of torsion is not yet ruled out from the experimental observations. The neutrino oscillation phenomenon can be used to probe the torsion, shown in \cite{mandal2021neutrino}. In the present article, we show the effect of torsion as a classical source on neutrino oscillation in Minkowski spacetime. Further, we show the change in the helicity state in neutrino oscillation due to the presence of torsion.

\section{Neutrino oscillation in the presence of a matter current}

Although neutrinos do not have an electric charge, there exists a conserved current corresponding to $U(1)$ symmetry. If neutrinos are propagating in a medium in the presence of a background conserved current, we expect an interaction between the background matter current and the propagating neutrinos. In this section, the effect of this interaction in neutrino oscillation is shown explicitly in a covariant manner. Here, we consider a phenomenological model in which the conserved matter current is coupled to the $U(1)$ current of neutrinos, which gives rise to a point interaction. This kind of interaction can in principle be obtained if a gauge field is minimally coupled to both the above-mentioned currents. 

\subsection{Dirac field theory in presence of a background conserved current}

The conserved current corresponding to $U(1)$ symmetry in Dirac field theory is given by $\bar{\psi}\gamma^{\mu}\psi$ where $\psi$ is the spinor field. A natural way to describe one such interaction between this current and the above-mentioned background source is by adding the Lorentz invariant term $J_{\mu}\bar{\psi}\gamma^{\mu}\psi$ in the Lagrangian density, where $J^{\mu}$ is the background conserved current. Considering the background conserved current as a function of $\phi=k.x$, the action can be expressed as 
\begin{equation}\label{reduced action}
S=\int d^{4}x\Big[i\bar{\psi}(x)\gamma^{\mu}\partial_{\mu}\psi(x)-m\bar{\psi}(x)\psi(x)-g J_{\mu}(\phi)\bar{\psi}(x)\gamma^{\mu}\psi(x)\Big],
\end{equation} 
where $g$ is the effective interaction coupling constant of mass dimension one and $k^{2}=0$. Since $J_{\mu}$ is covariantly conserved, $k_{\mu}\frac{dJ^{\mu}}{d\phi}=0$. Hence, apart from mass, there are two other dimensionful quantities which are coupling strength $g$ and the momentum $k$. The equation of motion corresponding to the action (\ref{reduced action}) is the following
\begin{equation}
(i\slashed{\partial}-g\slashed{J}(\phi)-m)\psi(x)=0.
\end{equation}
This is nothing but the Dirac-Volkov equation \cite{wolkow1935klasse}. The solution of above equation corresponding to a positive-energy particle carrying momentum $p$ is given by
\begin{equation}\label{solution1}
\psi_{p}(x)=\sqrt{\frac{m}{E}}\left(\mathbb{I}+g\frac{\slashed{k}\slashed{J}(\phi)}{k.p}\right)u(p) \ e^{-ip.x-i\int_{\phi_{0}}^{\phi}\Big[g\frac{p.J(\phi')}{k.p}-\frac{g^{2}J^{2}(\phi')}{2k.p}\Big]d\phi'},
\end{equation}
where $\sqrt{\frac{m}{E}}$ is a normalization factor, $E=p^{0}$. $u(p)$ is a spinor satisfying $(\slashed{p}-m)u(p)=0$, and $\bar{u}_{s}(p)\gamma^{\mu}u_{s'}(p)=\frac{p^{\mu}}{m}\delta_{ss'}$ where $p^{2}=m^{2}$. Here $s,s'=+,-$ denote helicity of states. On the other hand, the solution of anti-particle is given by
\begin{equation}\label{solution2}
\psi_{p}^{(anti)}(x)=\sqrt{\frac{m}{E}}\left(\mathbb{I}-g\frac{\slashed{k}\slashed{J}(\phi)}{k.p}\right)v(p) \ e^{-ip.x+i\int_{\phi_{0}}^{\phi}\Big[g\frac{p.J(\phi')}{k.p}+\frac{g^{2}J^{2}(\phi')}{2k.p}\Big]d\phi'},
\end{equation}
where $v(p)$ is a spinor satisfying $(\slashed{p}+m)v(p)=0$, and $\bar{v}_{s}(p)\gamma^{\mu}v_{s'}(p)=\frac{p^{\mu}}{m}\delta_{ss'}$ where $p^{2}=m^{2}$.

The neutrinos are electrically neutral, however, due to the global $U(1)$ symmetry, there is an associated conserved 4-current which is $\bar{\psi}\gamma^{\mu}\psi$, and the associated conserved charge is given by the number of neutrinos minus the number of its antiparticles.

Consider QED (Quantum Electrodynamics) in which we consider the electrons, protons and other charged particles couple to the photons. It is not hard to see that if we integrate out the photon degrees of freedom then there will be four-fermion interaction terms which involve different conserved 4-currents associated with different charged particles. This can also be seen in the two and higher loop Feynman diagrams at the level of generating functionals.

If an astrophysical source, for example supernova explosion of a star, is emitting neutrinos and they are coupled to the plasma ejected in some direction, then we expect naturally such a current-current interaction which breaks the Lorentz symmetry due to a preferred direction. This is the motivation behind considering the vector current interaction in this section. Moreover, the above-mentioned model can also be explained from the point view of induced electric charge of the neutrinos in a dispersive medium as shown in \cite{ORAEVSKY1987135}.

\subsection{Neutrino oscillation}

In mass eigenstates, the action describing neutrinos propagating in a background with a source current is given by
\begin{equation}
S=\int d^{4}x\sum_{j}\Big[i\bar{\psi}_{j}(x)\gamma^{\mu}\partial_{\mu}\psi_{j}(x)-m_{j}\bar{\psi}_{j}(x)\psi_{j}(x)-g J_{\mu}(\phi)\bar{\psi}_{j}(x)\gamma^{\mu}\psi_{j}(x)\Big], \ \phi=k.x,
\end{equation}
where $\{m_{j}\}$ are diagonal elements of diagonalized mass matrix. However, this mass matrix is non-diagonal in flavor eigenstates $\{\ket{\nu_{\alpha}}\}$ which are represented by linear superposition of the mass eigenstates $\{\ket{\nu_{i}}\}$:
\begin{equation}\label{trans}
\ket{\nu_{\alpha}}=\sum_{i}U_{\alpha i}\ket{\nu_{i}},
\end{equation}
where $U$ (PMNS matrix) is a unitary matrix through which we can transform mass matrix into a diagonal form and it is parametrized as
\begin{equation}
U=\begin{bmatrix}
c_{12}c_{13} & s_{12}c_{13} & s_{13}e^{-i\delta_{CP}}\\
-s_{12}c_{23} - c_{12}s_{23}s_{13}e^{i\delta_{CP}} & c_{12}c_{23} - s_{12}s_{23}s_{13}e^{i\delta_{CP}} & s_{23}c_{13}\\
s_{12}s_{23} - c_{12}c_{23}s_{13}e^{i\delta_{CP}} & - c_{12}s_{23} - s_{12}c_{23}s_{13}e^{i\delta_{CP}} & c_{23}c_{13},
\end{bmatrix}
\end{equation}
where $c_{ij}=\cos\theta_{ij}$ and $s_{ij}=\sin\theta_{ij}$. If a beam of pure $\nu_{\alpha}$ state is produced at initial time $t=0$, then the initial state is a superposition of the mass eigenstates
\begin{equation}
\ket{\nu_{\alpha}(0)}=\sum_{i}U_{\alpha i}\ket{\nu_{i}}.
\end{equation} 
The time evolution of a mass eigenstate $\ket{\nu_{i}}$ is determined by the Dirac equation for a propagating neutrino with definite mass $m_{i}$. From the Dirac equation, we obtain the following expression of time-evolved mass-eigenstate with momentum $\vec{p}$
\begin{equation}
\ket{\nu_{j},s}_{t}=\sqrt{\frac{m_{j}}{2E_{j}}}\left(\mathbb{I}+g\frac{\slashed{k}\slashed{J}(\phi)}{k.p_{j}}\right)u_{s}(p_{j}) \ e^{-iE_{j}t-i\int_{\phi_{0}}^{\phi}\Big[g\frac{p_{j}.J(\phi')}{k.p_{j}}-\frac{g^{2}J^{2}(\phi')}{2k.p_{j}}\Big]d\phi'}\ket{\nu_{j},s},
\end{equation}
where $p_{j}=(E_{j},\vec{p})$ and $E_{j}=\sqrt{\vec{p}^{2}+m_{j}^{2}}$. On the other hand, $\phi(t,\vec{x})\equiv\phi$ and $\phi(t=0,\vec{x})\equiv\phi_{0}$. As a consequence of (\ref{trans}), we obtain
\begin{equation}
\ket{\nu_{\alpha}(t),s}=\sum_{j}U_{\alpha j}\sqrt{\frac{m_{j}}{2E_{j}}}\left(\mathbb{I}+g\frac{\slashed{k}\slashed{J}(\phi)}{k.p_{j}}\right)u_{s}(p_{j}) \ e^{-iE_{j}t-i\int_{\phi_{0}}^{\phi}\Big[g\frac{p_{j}.J(\phi')}{k.p_{j}}-\frac{g^{2}J^{2}(\phi')}{2k.p_{j}}\Big]d\phi'}\ket{\nu_{j},s}.
\end{equation}
Therefore, the probability amplitude of observing an initially created flavor eigenstate $\ket{\nu_{\alpha}}$ as the flavor eigenstate $\ket{\nu_{\beta}}$ at some later time $t$ becomes
\begin{equation}\label{trans amp}
\begin{split}
\braket{\nu_{\beta},s_{2}|\nu_{\alpha}(t),s_{1}}=\sum_{j} & U_{\alpha j}U_{\beta j}^{*}\frac{m_{j}}{2E_{j}}\bar{u}_{s_{2}}(p_{j})\left(\mathbb{I}+g\frac{\slashed{J}(\phi_{0})\slashed{k}}{k.p_{j}}\right)\gamma^{0}\left(\mathbb{I}+g\frac{\slashed{k}\slashed{J}(\phi)}{k.p_{j}}\right)u_{s_{1}}(p_{j})\\
 & \times e^{-iE_{j}t-i\int_{\phi_{0}}^{\phi}\Big[g\frac{p_{j}.J(\phi')}{k.p_{j}}-\frac{g^{2}J^{2}(\phi')}{2k.p_{j}}\Big]d\phi'}.
\end{split}
\end{equation}
Hence, the probability for a transition $\nu_{\alpha}\rightarrow\nu_{\beta}$ under the time evolution is given by
\begin{equation}\label{trans prob}
\begin{split}
\mathcal{P}_{\nu_{\alpha,s_{1}}\rightarrow\nu_{\beta,s_{2}}}(t) & =|\braket{\nu_{\beta}|\nu_{\alpha}(t)}|^{2}=\sum_{j,l}U_{\alpha j}U_{\beta j}^{*}U_{\alpha l}^{*}U_{\beta l}e^{-i(E_{j}-E_{l})t}\mathcal{A}_{\vec{p}}(\phi,k,m_{j},t)\mathcal{A}_{\vec{p}}^{*}(\phi,k,m_{l},t)\\
 & \times e^{i\int_{\phi_{0}}^{\phi}\Big[g\frac{p_{l}.J(\phi')}{k.p_{l}}-\frac{g^{2}J^{2}(\phi')}{2k.p_{l}}\Big]d\phi'-i\int_{\phi_{0}}^{\phi}\Big[g\frac{p_{j}.J(\phi')}{k.p_{j}}-\frac{g^{2}J^{2}(\phi')}{2k.p_{j}}\Big]d\phi'},
\end{split}
\end{equation}
where $p_{l}=(E_{l},\vec{p}), \ E_{l}=\sqrt{\vec{p}^{2}+m_{l}^{2}}$, and
\begin{equation}
\mathcal{A}_{\vec{p};s_{1},s_{2}}(\phi,k,m_{j},t)\equiv\frac{m_{j}}{2E_{j}}\bar{u}_{s_{2}}(p_{j})\left(\mathbb{I}+g\frac{\slashed{J}(\phi_{0})\slashed{k}}{k.p_{j}}\right)\gamma^{0}\left(\mathbb{I}+g\frac{\slashed{k}\slashed{J}(\phi)}{k.p_{j}}\right)u_{s_{1}}(p_{j}).
\end{equation}
For high-energy neutrinos, $E_{j}-E_{l}$ in the exponent can be expressed as $E_{j}-E_{l}=\frac{m_{j}^{2}-m_{l}^{2}}{2|\vec{p}|}$. Note that for $g=0$, the expression $\mathcal{A}_{\vec{p};s_{1},s_{2}}(\phi,k,m_{j},t)$ reduces to $\delta_{s_{1}s_{2}}$. However, for $g \neq 0$, we expect the change in spin during the neutrino oscillation. On the other hand, in \cite{Dvornikov:2002rs}, it is claimed that vector current interaction does not play any role in the evolution of the spin of neutrinos during the neutrino oscillation. However, it is important to notice that evolution of spin operator in eq. (5) of \cite{Dvornikov:2002rs} indeed depends on the vector current interaction. However, under the approximation that the authors considered following \cite{Ternov:1990qp}, it is shown that the vector current interaction does not play any role on the evolution of the spin of neutrinos.

On the other hand, in our analysis, we have not made such kind of approximation, and the reason behind we expect the spin(-flavor) oscillations is due to the fact in the equation (\ref{trans amp}), $\slashed{J}(\phi_{0})\slashed{k}$ can be expressed as $k.J(\phi_{0}) + [\gamma^{\mu}, \gamma^{\nu}]k_{\mu}J_{\nu}(\phi_{0})$ where $[\gamma^{\mu}, \gamma^{\nu}]$ is related to spin operators for the spatial components $(\mu = i, \ \nu = j, \ i,j = \{1,2,3\})$. As a result of such a decomposition, we expect to have change in spin states during neutrino oscillations. The other reason behind such neutrino spin-flip due to the vector interaction could be because of the presence of the inhomogeneous current $J^{\mu}(\phi)$ \cite{Ansarifard:2024zxm}.

\section{Neutrino oscillation in the presence of a oscillating current}

Considering the background source current to be of the following form
\begin{equation}
J^{\mu}(x)=a^{\mu}\cos\phi, \ \phi=k.x,
\end{equation}
where $a^{\mu}$ is the dimensionless constant polarization vector of the current. Using the above expression in (\ref{solution1}), the positive-energy solution of the Dirac-Volkov equation becomes the following
\begin{equation}\label{solution oscillating current}
\psi_{p}(x)=\sqrt{\frac{m}{Q}}\left(1+g\frac{\slashed{k}\slashed{a}\cos\phi}{2k.p}\right)u(p)e^{-iq.x-ig\frac{a.p}{k.p}\sin\phi+i\frac{g^{2}a^{2}}{8k.p}\sin2\phi},
\end{equation}
where
\begin{equation}
q^{\mu}=p^{\mu}+\frac{g^{2}\bar{a}^{2}}{4k.p}k^{\mu}, \ \bar{a}=\sqrt{-a^{2}}.
\end{equation}
Since $k^{2}=0$, the four-vector $q^{\mu}$ satisfies the following relation
\begin{equation}
q^{2}=m^{2}+\frac{g^{2}\bar{a}^{2}}{2}\equiv m_{*}^{2}, \ a.q=a.p, \ k.p=k.q,
\end{equation}
where $m_{*}$ is the effective mass of the fermions and $Q=q^{0}$ in the equation (\ref{solution oscillating current}). Further, using the following result
\begin{equation}
e^{-i\alpha\sin\theta+i\beta\sin2\theta}=\sum_{s=-\infty}^{\infty}A_{0}(s,\alpha,\beta)e^{-is\theta},
\end{equation}
the solution can be expressed as follows
\begin{equation}
\psi_{p}(x)=\sqrt{\frac{m}{2Q}}\sum_{s=0}^{\infty}\left(A_{0}(s,\alpha,\beta)+g\frac{\slashed{k}\slashed{a}}{2k.p}A_{1}(s,\alpha,\beta)\right)u(p)e^{-i(q+sk).x},
\end{equation}
where $\alpha,\beta$ are identified with $g\frac{a.q}{k.q}$ and $\frac{g^{2}a^{2}}{8k.q}$ respectively. $A_{0}(s,\alpha,\beta)$ is the generalized Bessel function and $A_{1}(s,\alpha,\beta)$ is expressed as
\begin{equation}
A_{1}(s,\alpha,\beta)=\frac{1}{2}[A_{0}(s+1,\alpha,\beta)+A_{0}(s-1,\alpha,\beta)].
\end{equation}
Hence, we obtain the following expression for the transition amplitude between the flavor states
\begin{equation}\label{transition amp}
\begin{split}
\braket{\nu_{\beta},s_{2}|\nu_{\alpha}(t),s_{1}} & =\sum_{s=-\infty}^{\infty}\sum_{j}U_{\alpha j}U_{\beta j}^{*}\frac{m_{j}}{2Q_{j}}\bar{u}_{s_{2}}(p_{j})\left(A_{0}(s,\alpha_{j},\beta_{j})+g\frac{\slashed{a}\slashed{k}}{2k.p}A_{1}(s,\alpha_{j},\beta_{j})\right)\gamma^{0}\\
 & \times\left(A_{0}(s,\alpha_{j},\beta_{j})+g\frac{\slashed{k}\slashed{a}}{2k.p}A_{1}(s,\alpha_{j},\beta_{j})\right)u_{s_{1}}(p_{j})e^{-i(Q_{j}+s\omega)t},
\end{split}
\end{equation}
where $q_{j}^{\mu}=p_{j}^{\mu}+\frac{g^{2}\bar{a}^{2}}{4k.p_{j}}k^{\mu}, Q_{j}=q_{j}^{0}$ and $\omega=k^{0}$. Therefore, the transition probability becomes
\begin{equation}\label{transition prob}
\begin{split}
\mathcal{P}_{\nu_{\alpha},s_{1}\rightarrow\nu_{\beta},s_{2}} & (t,\vec{x})=|\braket{\nu_{\beta}|\nu_{\alpha}(t)}|^{2}=\sum_{n_{1},n_{2}=-\infty}^{\infty}
\sum_{j,l}\frac{m_{j}}{2Q_{j}}\frac{m_{l}}{2Q_{l}}U_{\alpha j}U_{\beta j}^{*}U_{\alpha l}^{*}U_{\beta l}e^{-i[(Q_{j}-Q_{l})+(n_{1}-n_{2})\omega]t}\\
\times\Bigg[\bar{u}_{s_{2}}(p_{j}) & \left(A_{0}(n_{1},\alpha_{j},\beta_{j})+g\frac{\slashed{a}\slashed{k}}{2k.p_{j}}A_{1}(n_{1},\alpha_{j},\beta_{j})\right)\gamma^{0}\\
 & \times\left(A_{0}(n_{1},\alpha_{j},\beta_{j})+g\frac{\slashed{k}\slashed{a}}{2k.p_{j}}A_{1}(n_{1},\alpha_{j},\beta_{j})\right)u_{s_{1}}(p_{j})\Bigg]\\
\times\Bigg[\bar{u}_{s_{2}}(p_{l}) & \left(A_{0}(n_{2},\alpha_{l},\beta_{l})+g\frac{\slashed{a}\slashed{k}}{2k.p_{l}}A_{1}(n_{2},\alpha_{l},\beta_{l})\right)\gamma^{0}\\
 & \times\left(A_{0}(n_{2},\alpha_{l},\beta_{l})+g\frac{\slashed{k}\slashed{a}}{2k.p_{l}}A_{1}(n_{2},\alpha_{l},\beta_{l})\right)u_{s_{1}}(p_{l})\Bigg]^{*}.
\end{split}
\end{equation}
The above two expressions of transition amplitude and probability of neutrino oscillation in (\ref{transition amp}) and (\ref{transition prob}), respectively show the nature of neutrino oscillation in the presence of background current. The expression of transition probability in (\ref{transition prob}) also shows that neutrino oscillation in this case also depends on $g$ and $\bar{a}^{2}$. In (\ref{transition amp}), the contribution comes from the states with an equal number of quanta with momentum $k$ present in the dressed fermionic states $\ket{\nu_{\alpha}(t),s_{1}}, \ket{\nu_{\beta},s_{2}}$, which is denoted by $s$. The concept of dressed fermionic states is often introduced in quantum electrodynamics in the presence of lasers \cite{lotstedt2009correlated}. Further, the expressions in (\ref{transition amp}) and (\ref{transition prob}) also imply that the oscillating background conserved current can change helicity states during neutrino oscillation. This is a generic feature of background conserved current which can be checked from (\ref{trans amp}) and (\ref{trans prob}).   

\section{Neutrino oscillation in a electromagnetic field}

\subsection{Dirac equation with a non-minimal coupling}

We consider the following action 
\begin{equation}
S=\int d^{4}x\Big[\bar{\psi}(i\slashed{\partial}-m)\psi-i\frac{\mu}{2}F_{\mu\nu}\bar{\psi}[\gamma^{\mu},\gamma^{\nu}]\psi\Big],
\end{equation}
where the second term describes a non-minimal coupling between background electromagnetic gauge field and fermions. We will come to the coupling $\mu$ later. The corresponding equation of motion is given by
\begin{equation}
\Big[(i\slashed{\partial}-m)-i\mu F_{\mu\nu}\gamma^{\mu}\gamma^{\nu}\Big]\psi=0.
\end{equation}
Like the previous section, we consider $A_{\mu}(x)=A_{\mu}(\phi)=\varepsilon_{\mu}f(\phi)$ where $\phi=k.x$ and $k^{2}=0, \ k.A(\phi)=0$. This makes the above equation
\begin{equation}
\Big[(i\slashed{\partial}-m)-i\mu [\slashed{k},\slashed{A}'(\phi)]\Big]\psi=0,
\end{equation}
where $A'_{\mu}(\phi)=\frac{dA_{\mu}}{d\phi}=\varepsilon_{\mu}\frac{df}{d\phi}$. In order to solve the above equation, we can multiply the above equation with the operator $\Big[(i\slashed{\partial}+m)-i\mu [\slashed{k},\slashed{A}'(\phi)]\Big]$ from the left and we obtain the following relation
\begin{equation}\label{Dirac eqn with multiplication}
\begin{split}
\Big[-\partial^{2}-m^{2} & +\mu\slashed{\partial}[\slashed{k},\slashed{A}'(\phi)]+\mu[\slashed{k},\slashed{A}'(\phi)]\slashed{\partial}-\mu^{2}[\slashed{k},\slashed{A}'(\phi)]^{2}\Big]\psi=0.
\end{split}
\end{equation}
We consider the ansatz of the following form
\begin{equation}
\psi_{p}(x)=e^{-ip.x}F_{p}(\phi)u(p),
\end{equation}
which makes the equation (\ref{Dirac eqn with multiplication}) following 
\begin{equation}\label{Dirac equation 2}
\begin{split}
\Big[-2k.pF_{p}'(\phi)+\mu\slashed{p}[\slashed{k},\slashed{A}'(\phi)]F_{p}(\phi) & +\mu[\slashed{k},\slashed{A}'(\phi)]\slashed{p}F_{p}(\phi)\Big]u(p)=0,
\end{split}
\end{equation}
where we used the following identities and results
\begin{equation}
\begin{split}
\gamma^{\mu}\gamma^{\nu}\gamma^{\rho} & =\gamma^{\mu}\eta^{\nu\rho}+\gamma^{\rho}\eta^{\mu\nu}-\gamma^{\nu}\eta^{\mu\rho}-i\epsilon^{\sigma\mu\nu\rho}\gamma_{\sigma}\gamma^{5}\\
\implies[\slashed{k},\slashed{A}'(\phi)]\slashed{k} & =\slashed{k}[\slashed{k},\slashed{A}'(\phi)]=\slashed{k}[\slashed{k},\slashed{A}''(\phi)]=0=[\slashed{k},\slashed{A}'(\phi)]^{2}.
\end{split}
\end{equation}
Further, the on-shell relation is $p^{2}=m^{2}$ and the positive-energy spinor satisfies $(\slashed{p}-m)u(p)=0$. On the other hand, the solution of the equation (\ref{Dirac equation 2})
\begin{equation}
2k.pF_{p}'(\phi)=\mu\slashed{p}[\slashed{k},\slashed{A}'(\phi)]F_{p}(\phi)+\mu[\slashed{k},\slashed{A}'(\phi)]\slashed{p}F_{p}(\phi),
\end{equation}
is given by
\begin{equation}
F_{p}(\phi)=e^{\frac{\mu}{2k.p}\int_{\phi_{0}}^{\phi}\{\slashed{p},[\slashed{k},\slashed{A}'(\phi')]\}d\phi'}.
\end{equation}
Hence, the positive-energy solution of Dirac fermion is given by
\begin{equation}\label{Dirac magnetic moment}
\psi_{p}(x)=\sqrt{\frac{m}{2E}}e^{-ip.x+\frac{\mu}{2k.p}\int_{\phi_{0}}^{\phi}\{\slashed{p},[\slashed{k},\slashed{A}'(\phi')]\}d\phi'}u(p),
\end{equation}
where
\begin{equation}
\{\slashed{p},[\slashed{k},\slashed{A}'(\phi)]\}=2p_{\mu}k_{\nu}A'_{\rho}(\phi)[\gamma^{\mu}\gamma^{\nu}\gamma^{\rho}+\gamma^{\nu}\gamma^{\rho}\gamma^{\mu}]=-4ip_{\mu}k_{\nu}A'_{\rho}(\phi)\epsilon^{\sigma\nu\rho\mu}\gamma_{\sigma}\gamma^{5}.
\end{equation}
Therefore, the solution in (\ref{Dirac magnetic moment}) can be expressed as
\begin{equation}
\psi_{p}(x)=\sqrt{\frac{m}{2E}}e^{-ip.x-i\frac{2\mu}{k.p}\int_{\phi_{0}}^{\phi}p_{\mu}k_{\nu}A'_{\rho}(\phi')\epsilon^{\sigma\nu\rho\mu}\gamma_{\sigma}\gamma^{5}d\phi'}u(p).
\end{equation}

\subsection{Neutrino oscillation in a background electromagnetic field}

Although neutrinos are chargeless fermions, it is proposed that the neutrinos carry non-zero magnetic moment due to their masses which can couple to electromagnetic gauge field \cite{cisneros1971effect, fujikawa1980magnetic}. An upper bound on the values of these magnetic moments is also put in \cite{bell2006model, akhmedov2003solar, canas2016updating, beacom1999neutrino}. Considering the non-minimal model suggested in \cite{giunti2009neutrino, jenkins2013gauge}, we consider the following action describing the interaction between neutrinos and electromagnetic gauge fields
\begin{equation}
S=\int d^{4}x\Big[\sum_{i}\bar{\psi}_{i}(i\slashed{\partial}-m_{i})\psi_{i}-i\sum_{jk}\frac{\mu_{jk}}{2}F_{\mu\nu}\bar{\psi}_{j}[\gamma^{\mu},\gamma^{\nu}]\psi_{k}\Big],
\end{equation}
where $\mu_{jk}$ is a non-diagonal magnetic moment matrix in the mass-eigenstates basis of neutrinos and $A_{\mu}(x)=A_{\mu}(\phi)=\varepsilon_{\mu}f(\phi)$ with $\phi=k.x, k^{2}=0$, and $k.A(\phi)=0$. The expression for $\mu_{jk}$ derived in \cite{balantekin2014magnetic, broggini2012electromagnetic} shows that the non-diagonal elements of the magnetic moment matrix are small compared to the diagonal terms, hence, in leading order, we can write the above action as
\begin{equation}
S=\int d^{4}x\sum_{j}\Big[\bar{\psi}_{j}(i\slashed{\partial}-m_{i})\psi_{j}-i\frac{\mu_{j}}{2}F_{\mu\nu}\bar{\psi}_{j}[\gamma^{\mu},\gamma^{\nu}]\psi_{j}\Big],
\end{equation}
where $\mu_{j}\equiv\mu_{jj}$. Therefore, using the expression in (\ref{Dirac magnetic moment}), we obtain the following expression for the transition amplitude of neutrino oscillation
\begin{equation}
\begin{split}
\braket{\nu_{\beta}(t'),s_{2}|\nu_{\alpha}(t),s_{1}} & = \sum_{j}\frac{m_{j}}{2E_{j}}U_{\alpha j}U_{\beta j}^{*} \bar{u}_{s_{2}}(p_{j})e^{-i\frac{2\mu_{j}}{k.p_{j}}\int_{\phi_{0}}^{\phi'}p_{j,\mu}k_{\nu}A'_{\rho}(\bar{\phi})\epsilon^{\sigma\nu\rho\mu}\gamma^{5}\gamma_{\sigma}d\bar{\phi}}\gamma^{0}\\
 & \times e^{-i\frac{2\mu_{j}}{k.p_{j}}\int_{\phi_{0}}^{\phi}p_{j,\mu}k_{\nu}A'_{\rho}(\bar{\phi})\epsilon^{\sigma\nu\rho\mu}\gamma_{\sigma}\gamma^{5}d\bar{\phi}}u_{s_{1}}(p_{j})e^{-iE_{j}(t-t')},
\end{split}
\end{equation}
where $\phi'=k.x'$ and $x'^{0}=t', \ x'^{i}=x^{i}$. Hence, the expression for the corresponding transition probability is given by
\begin{equation}
\begin{split}
\mathcal{P}_{\nu_{\alpha},s_{1}\rightarrow\nu_{\beta},s_{2}}(t,t') & =|\braket{\nu_{\beta}(t'),s_{2}|\nu_{\alpha}(t),s_{1}}|^{2}=\sum_{j,l}\frac{m_{j}}{2E_{j}}\frac{m_{l}}{2E_{l}}U_{\alpha j}U_{\beta j}^{*}U_{\alpha l}^{*}U_{\beta l}e^{-i(E_{j}-E_{l})(t-t')}\\
\times\Bigg[\bar{u}_{s_{2}}(p_{j}) & e^{-i\frac{2\mu_{j}}{k.p_{j}}\int_{\phi_{0}}^{\phi'}p_{j,\mu}k_{\nu}A'_{\rho}(\bar{\phi})\epsilon^{\sigma\nu\rho\mu}\gamma^{5}\gamma_{\sigma}d\bar{\phi}}\gamma^{0}e^{-i\frac{2\mu_{j}}{k.p_{j}}\int_{\phi_{0}}^{\phi}p_{j,\mu}k_{\nu}A'_{\rho}(\bar{\phi})\epsilon^{\sigma\nu\rho\mu}\gamma_{\sigma}\gamma^{5}d\bar{\phi}}u_{s_{1}}(p_{j})\Bigg]\\
\times\Bigg[\bar{u}_{s_{2}}(p_{l}) & e^{-i\frac{2\mu_{l}}{k.p_{l}}\int_{\phi_{0}}^{\phi'}p_{l,\mu}k_{\nu}A'_{\rho}(\bar{\phi})\epsilon^{\sigma\nu\rho\mu}\gamma^{5}\gamma_{\sigma}d\bar{\phi}}\gamma^{0}e^{-i\frac{2\mu_{l}}{k.p_{l}}\int_{\phi_{0}}^{\phi}p_{l,\mu}k_{\nu}A'_{\rho}(\bar{\phi})\epsilon^{\sigma\nu\rho\mu}\gamma_{\sigma}\gamma^{5}d\bar{\phi}}u_{s_{1}}(p_{l})\Bigg]^{*}.
\end{split}
\end{equation}
The above expression shows that both the amplitude and nature of oscillation depend on the magnetic moments of neutrinos, four-momenta $k^{\mu}$, and the derivative of the gauge field. Further, it also shows that in general, helicity changes in neutrino oscillation due to the coupling between magnetic moments of neutrinos and electromagnetic gauge field. The above conclusions are indeed consistent with the results found in \cite{Dvornikov:2018tmm, Dvornikov:2019pxd}. However, the approach that we followed here is completely different compared to the one followed in \cite{Dvornikov:2018tmm, Dvornikov:2019pxd}. Moreover, in our analysis, we considered the magnetic moments of neutrinos to be different in mass-basis.

\section{Effect of torsion in neutrino oscillation in Minkowski spacetime}

Following the results in \cite{mandal2021neutrino}, the Dirac equation in Minkowski spacetime in the presence of torsion can be expressed as
\begin{equation}
\Big[i\slashed{\partial}-m+\frac{i}{8}K_{\rho\sigma\mu}\gamma^{\mu}[\gamma^{\rho},\gamma^{\sigma}]\Big]\psi=0.
\end{equation}
where $K_{\rho\sigma\mu}=-K_{\sigma\rho\mu}$ is contorsion tensor. Here, we restrict our discussion to the case in which $K_{\rho\sigma\mu}(x)=\kappa_{\rho\sigma\mu}K(\phi)$ is a plane-wave field where $\phi=k.x, \ k^{2}=0$. Now we act the operator $\Big[i\slashed{\partial}+m+\frac{i}{8}K_{\rho\sigma\mu}\gamma^{\mu}[\gamma^{\rho},\gamma^{\sigma}]\Big]$ on the above equation from the left and obtain the following relation
\begin{equation}
\begin{split}
\Big[-\partial^{2} & -m^{2}-\frac{1}{8}\slashed{\partial}K_{\rho\sigma\mu}\gamma^{\mu}[\gamma^{\rho},\gamma^{\sigma}]-\frac{1}{8}K_{\rho\sigma\mu}\gamma^{\mu}[\gamma^{\rho},\gamma^{\sigma}]\slashed{\partial}-\frac{1}{64}\left(K_{\rho\sigma\mu}\gamma^{\mu}[\gamma^{\rho},\gamma^{\sigma}]\right)^{2}\Big]\psi=0.
\end{split}
\end{equation} 
Using the ansatz $\psi_{p}(x)=e^{-ip.x}\mathcal{G}(\phi)u(p)$, the above equation becomes
\begin{equation}\label{Dirac equation torsion}
\begin{split}
\Big[p^{2}\mathcal{G}(\phi)+2ik.p\mathcal{G}'(\phi) & -m^{2}\mathcal{G}(\phi)+\frac{i}{4}\slashed{p}K_{\rho\sigma\mu}\gamma^{\mu}\gamma^{\rho}\gamma^{\sigma}\mathcal{G}(\phi)+\frac{i}{4}K_{\rho\sigma\mu}\gamma^{\mu}\gamma^{\rho}\gamma^{\sigma}\slashed{p}\mathcal{G}(\phi)\\
-\frac{1}{4}\slashed{k}K'_{\rho\sigma\mu}\gamma^{\mu}\gamma^{\rho}\gamma^{\sigma}\mathcal{G}(\phi) & -\frac{1}{4}\slashed{k}K_{\rho\sigma\mu}\gamma^{\mu}\gamma^{\rho}\gamma^{\sigma}\mathcal{G}'(\phi)-\frac{1}{4}K_{\rho\sigma\mu}\gamma^{\mu}\gamma^{\rho}\gamma^{\sigma}\slashed{k}\mathcal{G}'(\phi)\\
 & -\frac{1}{16}\left(K_{\rho\sigma\mu}\gamma^{\mu}\gamma^{\rho}\gamma^{\sigma}\right)^{2}\mathcal{G}(\phi)\Big]u(p)=0.
\end{split}
\end{equation}
We define the following quantity
\begin{equation}
\mathcal{K}(\phi)\equiv K_{\rho\sigma\mu}(\phi)\gamma^{\mu}\gamma^{\rho}\gamma^{\sigma}=K_{\rho\sigma\mu}(\phi)[2\gamma^{\sigma}\eta^{\mu\rho}-i\epsilon^{\lambda\mu\rho\sigma}\gamma_{\lambda}\gamma^{5}],
\end{equation}
in terms of which the equation (\ref{Dirac equation torsion}) can be expressed as
\begin{equation}
k.p\mathcal{G}'(\phi)+\frac{1}{8}\{\slashed{p},\mathcal{K}\}\mathcal{G}(\phi)+\frac{i}{8}\slashed{k}\mathcal{K}'\mathcal{G}(\phi)+\frac{i}{8}\{\slashed{k},\mathcal{K}\}\mathcal{G}'(\phi)+\frac{i}{16}\mathcal{K}^{2}\mathcal{G}(\phi)=0.
\end{equation}
Solution of the above equation is given by
\begin{equation}
\mathcal{G}(\phi)=e^{\int_{\phi_{0}}^{\phi}\Big[k.p\mathbb{I}+\frac{i}{8}\{\slashed{k},\mathcal{K}(\bar{\phi})\}\Big]^{-1}\Big[-\frac{1}{8}\{\slashed{p},\mathcal{K}(\bar{\phi})\}-\frac{i}{8}\slashed{k}\mathcal{K}'(\bar{\phi})-\frac{i}{16}\mathcal{K}^{2}(\bar{\phi})\Big]d\bar{\phi}},
\end{equation}
with the boundary condition $\mathcal{G}(\phi_{0})=\mathbb{I}$. Hence, the positive-energy solution is given by
\begin{equation}
\psi_{p}(x)=\sqrt{\frac{m}{2E}}e^{-ip.x-\frac{1}{8}\int_{\phi_{0}}^{\phi}\Big[k.p\mathbb{I}+\frac{i}{8}\{\slashed{k},\mathcal{K}(\bar{\phi})\}\Big]^{-1}\Big[\{\slashed{p},\mathcal{K}(\bar{\phi})\}+i\slashed{k}\mathcal{K}'(\bar{\phi})+\frac{i}{2}\mathcal{K}^{2}(\bar{\phi})\Big]d\bar{\phi}}u(p)\equiv e^{-ip.x}\mathcal{M}_{p}(\phi)u_{p}.
\end{equation}
Therefore, the expression of transition amplitude for the neutrino oscillation in this case is given by
\begin{equation}
\braket{\nu_{\beta},s_{2}|\nu_{\alpha}(t),s_{1}}=\sum_{j}\frac{m_{j}}{2E_{j}}U_{\alpha j}U_{\beta j}^{*}\bar{u}_{s_{2}}(p_{j})\gamma^{0}\mathcal{M}_{p_{j}}(\phi)u_{s_{1}}(p_{j})e^{-iE_{j}t},
\end{equation}
whereas the transition probability becomes
\begin{equation}
\begin{split}
\mathcal{P}_{\nu_{\alpha},s_{1}\rightarrow\nu_{\beta},s_{2}}(t) & =|\braket{\nu_{\beta},s_{2}|\nu_{\alpha}(t),s_{1}}|^{2}=\sum_{j,l}U_{\alpha j}U_{\beta j}^{*}U_{\alpha l}^{*}U_{\beta l}e^{-i(E_{j}-E_{l})t}\Big[\\
 & [\bar{u}_{s_{2}}(p_{j})\gamma^{0}\mathcal{M}_{p_{j}}(\phi)u_{s_{1}}(p_{j})]\times[\bar{u}_{s_{2}}(p_{l})\gamma^{0}\mathcal{M}_{p_{l}}(\phi)u_{s_{1}}(p_{l})]^{*}\Big]\frac{m_{j}}{2E_{j}}\frac{m_{l}}{2E_{l}}.
\end{split}
\end{equation}
The above expression also shows that non-trivial torsion can indeed change the helicity state in neutrino oscillation.

\section{Neutrino oscillation in the presence of gravitational waves}

The action describing the Dirac-fermions interacting with GW is given by
\begin{equation}\label{GW0}
S[\bar{\psi},\psi]=\int d^{4}x\Big[\bar{\psi}(i\slashed{\partial}-m)\psi+i\kappa h_{\mu\nu}\bar{\psi}\gamma^{\mu}\partial^{\nu}\psi\Big],
\end{equation}
where $\kappa=\sqrt{8\pi G}$, and we have used the stress-energy tensor $T^{\mu\nu}(x)=\frac{i}{2}\bar{\psi}(x)[\gamma^{(\mu}\partial^{\nu)}-\gamma^{(\mu}\overleftarrow{\partial}^{\nu)}]\psi(x)$ of the Dirac-fermions. The corresponding equation of motion is given by
\begin{equation}\label{GW1}
(i\slashed{\partial}-m)\psi(x)+i\kappa h_{\mu\nu}(x)\gamma^{\mu}\partial^{\nu}\psi(x)=0.
\end{equation}
Like earlier, we consider the gravitational wave to be a plane-wave field of the following form
\begin{equation}
h_{\mu\nu}(x)=\epsilon_{\mu\nu}\mathcal{R}[\phi], \ \epsilon_{ \ \mu}^{\mu}=0, \ \phi=k.x, \ k^{2}=0,
\end{equation}
where $\epsilon_{\mu\nu}$ denotes the polarization of the gravitational wave and it satisfies $k^{\mu}\epsilon_{\mu\nu}=0$ which follows from the gauge choice $\partial^{\mu}h_{\mu\nu}=0$. Multiplying the equation (\ref{GW1}) by the operator $(i\slashed{\partial}+m)+i\kappa h_{\mu\nu}(x)\gamma^{\mu}\partial^{\nu}$ from the left, we obtain the following equation
\begin{equation}\label{GW2}
\begin{split}
\Big[-\Box & -m^{2}-\kappa\slashed{\partial}h_{\mu\nu}(x)\gamma^{\mu}\partial^{\nu}-\kappa h_{\mu\nu}(x)\gamma^{\mu}\partial^{\nu}\slashed{\partial}-\kappa^{2}h_{\mu\nu}(x)\gamma^{\mu}\partial^{\nu}h_{\rho\sigma}(x)\gamma^{\rho}\partial^{\sigma}\Big]\psi(x)=0,
\end{split}
\end{equation}
where every $\partial$ operator acts on all the quantities in its right. We consider the ansatz $\psi_{p}(x)=\mathcal{O}_{p}[\phi]u(p)e^{-ip.x}$ for the positive energy solution of the above equation (\ref{GW2}). Plugging this ansatz into the equation (\ref{GW2}), we obtain the following relation
\begin{equation}\label{GW3}
\begin{split}
\Big[(p^{2} & -m^{2})\mathcal{O}_{p}[\phi]+2ik.p\mathcal{O}'_{p}[\phi]+\kappa
\epsilon_{\mu\nu}\mathcal{R}[\phi]\slashed{p}\left(\gamma^{\mu}p^{\nu}\mathcal{O}_{p}[\phi]+i\gamma^{\mu}k^{\nu}\mathcal{O}'_{p}[\phi]\right)\\
 & +\kappa\epsilon_{\mu\nu}\mathcal{R}[\phi]\left(\gamma^{\mu}p^{\nu}\slashed{p}\mathcal{O}_{p}[\phi]+i\gamma^{\mu}k^{\nu}\slashed{p}\mathcal{O}'_{p}[\phi]\right)\\
 & +i\kappa\epsilon_{\mu\nu}\mathcal{R}'[\phi]\slashed{k}\left(\gamma^{\mu}p^{\nu}\mathcal{O}_{p}[\phi]+i\gamma^{\mu}k^{\nu}\mathcal{O}'_{p}[\phi]\right)\\
 & +i\kappa\epsilon_{\mu\nu}\mathcal{R}[\phi]\slashed{k}\left(\gamma^{\mu}p^{\nu}\mathcal{O}'_{p}[\phi]+i\gamma^{\mu}k^{\nu}\mathcal{O}''_{p}[\phi]\right)\\
 & +i\kappa\epsilon_{\mu\nu}\mathcal{R}[\phi]\slashed{k}\left(\gamma^{\mu}p^{\nu}\mathcal{O}_{p}[\phi]+i\gamma^{\mu}k^{\nu}\mathcal{O}'_{p}[\phi]\right)\mathcal{O}'_{p}[\phi]\\
 & +i\kappa\epsilon_{\mu\nu}\mathcal{R}[\phi]\left(\gamma^{\mu}p^{\nu}\slashed{k}\mathcal{O}_{p}[\phi]+i\gamma^{\mu}k^{\nu}\slashed{k}\mathcal{O}'_{p}[\phi]\right)\mathcal{O}'_{p}[\phi]\\
 & +i\kappa^{2}\mathcal{R}[\phi]\mathcal{R}'[\phi]\epsilon_{\mu\nu}\epsilon_{\rho\sigma}\gamma^{\mu}k^{\nu}\left(\gamma^{\rho}p^{\sigma}\mathcal{O}_{p}[\phi]+i\gamma^{\rho}k^{\sigma}\mathcal{O}'_{p}[\phi]\right)\\
 & +i\kappa^{2}\mathcal{R}^{2}[\phi]\epsilon_{\mu\nu}\epsilon_{\rho\sigma}\gamma^{\mu}k^{\nu}\left(\gamma^{\rho}p^{\sigma}\mathcal{O}'_{p}[\phi]+i\gamma^{\rho}k^{\sigma}\mathcal{O}''_{p}[\phi]\right)\\
 & +\kappa^{2}\epsilon_{\mu\nu}\epsilon_{\rho\sigma}\mathcal{R}^{2}[\phi]\Big[\gamma^{\mu}p^{\nu}\gamma^{\rho}p^{\sigma}\mathcal{O}_{p}^{2}[\phi]+i\gamma^{\mu}k^{\nu}\gamma^{\rho}p^{\sigma}\mathcal{O}_{p}[\phi]\mathcal{O}'_{p}[\phi]\\
 & +i\gamma^{\mu}p^{\nu}\gamma^{\rho}k^{\sigma}\mathcal{O}_{p}[\phi]\mathcal{O}'_{p}[\phi]-\gamma^{\mu}k^{\nu}\gamma^{\rho}k^{\sigma}\mathcal{O}_{p}^{'2}[\phi]\Big]\Big]=0.
\end{split}
\end{equation}
The transverse traceless condition reduces the above equation (\ref{GW3}) in the following form
\begin{equation}
\begin{split}
\Big[2ik.p\mathcal{O}'_{p}[\phi] & +\kappa\epsilon_{\mu\nu}\mathcal{R}[\phi]\slashed{p}\gamma^{\mu}p^{\nu}\mathcal{O}_{p}[\phi]+\kappa\epsilon_{\mu\nu}\mathcal{R}[\phi]\gamma^{\mu}p^{\nu}\slashed{p}\mathcal{O}_{p}[\phi]\\
 & +i\kappa\epsilon_{\mu\nu}\mathcal{R}'[\phi]\slashed{k}\gamma^{\mu}p^{\nu}\mathcal{O}_{p}[\phi]
+i\kappa\epsilon_{\mu\nu}\mathcal{R}[\phi]\slashed{k}\gamma^{\mu}p^{\nu}\mathcal{O}'_{p}[\phi]\\
 & +i\kappa\epsilon_{\mu\nu}\mathcal{R}[\phi]\slashed{k}\gamma^{\mu}p^{\nu}\mathcal{O}_{p}[\phi]\mathcal{O}'_{p}[\phi]+i\kappa\epsilon_{\mu\nu}\mathcal{R}[\phi]\gamma^{\mu}\slashed{k}p^{\nu}\mathcal{O}_{p}[\phi]\mathcal{O}'_{p}[\phi]\\
 & +\kappa^{2}\epsilon_{\mu\nu}\epsilon_{\rho\sigma}\mathcal{R}^{2}[\phi]\gamma^{\mu}p^{\nu}\gamma^{\rho}p^{\sigma}\mathcal{O}_{p}^{2}[\phi]\Big]=0,
\end{split}
\end{equation}
where we have used the dispersion $p^{2}=m^{2}$. Using the identity $\epsilon_{\mu\nu}\{\slashed{k},\gamma^{\mu}\}=2\epsilon_{\mu\nu}k^{\mu}=0$ in the above equation, we obtain the following first-order differential equation in $\mathcal{O}_{p}[\phi]$
\begin{align}\label{GW4}
\Big[2ik.p+i\kappa\epsilon_{\mu\nu}\mathcal{R}[\phi]\slashed{k}\gamma^{\mu}p^{\nu}\Big]\mathcal{O}'_{p}[\phi] & = -[2\kappa\epsilon_{\mu\nu}\mathcal{R}[\phi]p^{\mu}p^{\nu}+i\kappa\epsilon_{\mu\nu} \mathcal{R}'[\phi]\slashed{k}\gamma^{\mu}p^{\nu}]\mathcal{O}_{p}[\phi]\nonumber\\
 & -\kappa^{2}\epsilon_{\mu\nu}\epsilon_{ \ \sigma}^{\mu}\mathcal{R}^{2}[\phi]p^{\nu}p^{\sigma}\mathcal{O}_{p}^{2}[\phi]. 
\end{align}
Since $\kappa^{2}\ll1$, we neglect it in the above equation. Further, in order to consider $\kappa^{2}$ term in the above equation, we also need to consider $\mathcal{O}(\kappa^{2})$ terms in the action (\ref{GW0}) which generates $\mathcal{O}(\kappa^{3})$ term in the above equation, then we need to consider $\mathcal{O}(\kappa^{3})$ terms in the action (\ref{GW0}) and it goes on. Therefore, we truncate the terms beyond $\mathcal{O}(\kappa)$. Hence, the equation (\ref{GW4}) reduces to the following 
\begin{equation}
\mathcal{O}'_{p}[\phi]=i\Big[2k.p+\kappa\epsilon_{\mu\nu}\mathcal{R}[\phi]\slashed{k}\gamma^{\mu}p^{\nu}\Big]^{-1}\left(2\kappa\epsilon_{\mu\nu}\mathcal{R}[\phi]p^{\mu}p^{\nu}+i\kappa\epsilon_{\mu\nu}\mathcal{R}'[\phi]\slashed{k}\gamma^{\mu}p^{\nu}\right)\mathcal{O}_{p}[\phi]\equiv iU[p,k;\phi]\mathcal{O}_{p}[\phi].
\end{equation}
Solution of the above equation is given by
\begin{equation}
\mathcal{O}_{p}[\phi]=e^{i\int_{\phi_{0}}^{\phi}U[p,k;\bar{\phi}]d\bar{\phi}}\mathcal{O}_{p}[\phi_{0}].
\end{equation}
Hence, the positive energy solution of the Dirac equation (\ref{GW1}) becomes
\begin{equation}
\psi_{p}(x)=\sqrt{\frac{m}{2E_{p}}}e^{i\int_{\phi_{0}}^{\phi}U[p,k;\bar{\phi}]d\bar{\phi}}\mathcal{O}_{p}[\phi_{0}]u(p)e^{i\vec{p}.\vec{x}-iE_{p}t}.
\end{equation}
Therefore, the expression for transition amplitude in this case is given by
\begin{equation}\label{transition amp in GW}
\braket{\nu_{\beta},s_{2}|\nu_{\alpha}(t),s_{1}}=\sum_{j}\frac{m_{j}}{2E_{j}}U_{\alpha j}U_{\beta j}^{*}\bar{u}_{s_{2}}(p_{j})\gamma^{0}e^{i\int_{\phi_{0}}^{\phi}U[p_{j},k;\bar{\phi}]d\bar{\phi}}u_{s_{1}}(p_{j})e^{-iE_{j}t},
\end{equation}
where we consider $\mathcal{O}_{p}[\phi_{0}]=\mathbb{I}$. Hence, the transition probability becomes
\begin{equation}
\begin{split}
\mathcal{P}_{\nu_{\alpha},s_{1}\rightarrow\nu_{\beta},s_{2}}(t) & =|\braket{\nu_{\beta},s_{2}|\nu_{\alpha}(t),s_{1}}|^{2}=\sum_{j,l}U_{\alpha j}U_{\beta j}^{*}U_{\alpha l}^{*}U_{\beta l}e^{-i(E_{j}-E_{l})t}\Big[\\
 & \left(\bar{u}_{s_{2}}(p_{j})\gamma^{0}e^{i\int_{\phi_{0}}^{\phi}U[p_{j},k;\bar{\phi}]d\bar{\phi}}u_{s_{1}}(p_{j})\right)\\
 & \times\left(\bar{u}_{s_{1}}(p_{l})\gamma^{0}e^{-i\int_{\phi_{0}}^{\phi}U^{\dagger}[p_{l},k;\bar{\phi}]d\bar{\phi}}u_{s_{2}}(p_{l})\right)\Big]\frac{m_{j}}{2E_{j}}\frac{m_{l}}{2E_{l}}.
\end{split}
\end{equation}
The non-trivial nature of the operator $U[p,k;\phi]$ in the exponential changes the spin state of neutrinos during the neutrino oscillation. This follows from the fact that $U[p,k;\bar{\phi}]$ contains terms like $\epsilon_{\mu\nu}\mathcal{R}[\phi]\slashed{k}\gamma^{\mu}p^{\nu}$ and $\epsilon_{\mu\nu}\mathcal{R}'[\phi]\slashed{k}\gamma^{\mu}p^{\nu}$ which can be expressed as
\begin{equation}
\begin{split}
\epsilon_{\mu\nu} & \mathcal{R}[\phi]\slashed{k}\gamma^{\mu}p^{\nu} = \epsilon_{\mu\nu}\mathcal{R}[\phi]k^{\mu}p^{\nu} + \frac{1}{2}\epsilon_{\mu\nu}\mathcal{R}[\phi]k_{\rho}[\gamma^{\rho},\gamma^{\mu}]p^{\nu} = \frac{1}{2}\epsilon_{\mu\nu}\mathcal{R}[\phi]k_{\rho}[\gamma^{\rho},\gamma^{\mu}]p^{\nu}\\
\epsilon_{\mu\nu} & \mathcal{R}'[\phi]\slashed{k}\gamma^{\mu}p^{\nu} = \frac{1}{2}\epsilon_{\mu\nu}\mathcal{R}'[\phi]k_{\rho}[\gamma^{\rho},\gamma^{\mu}]p^{\nu} 
\end{split}
\end{equation}
where we used the transversality condition $\epsilon_{\mu\nu}k^{\mu} = 0$. In the above expressions, we can see the appearance of the spin operators for spatial components of $\rho$ and $\mu$. It is also important here to emphasise that the computation of transition amplitude in equation (\ref{transition amp in GW}) involves the initial state which effectively evolves under the exponential operator $e^{i\int_{\phi_{0}}^{\phi}U[p_{j},k;\bar{\phi}]d\bar{\phi}}$ in the Schr$\ddot{\text{o}}$dinger picture. This operator in a way captures the essence of one-particle scattering in the presence of gravitons which transfers non-zero four-momentum $k$.
Thus, GW affects the nature of neutrino oscillation, although this effect is relatively small since the coupling $\kappa$ is very small.

\section{Discussion}

Neutrinos play an important role in understanding various astrophysical and cosmological phenomena \cite{dolgov2002neutrinos, abazajian2017sterile, boyarsky2009role, dupuy2014describing}. In order to predict and explain astrophysical and cosmological observations through models, it is often required to know about the fluxes of different flavors of neutrinos. Hence, the nature of neutrino oscillations in the presence of background classical sources must be understood. Here, using the Dirac-Volkov equation, we show explicitly the nature of neutrino oscillations in the presence of a class of plane-wave classical sources. Further, the effect of coupling between the magnetic moments of neutrinos and the electromagnetic gauge field is also shown explicitly. Similarly, we also show that the couplings with torsion and GW in Minkowski spacetime also affect the nature of neutrino oscillation in which the helicity state can change. This is consistent with the results in the literature \cite{cirilo2013solar, cirilo2019fermion, dvornikov2020spin, lambiase2005neutrino, dvornikov2019neutrino} discussed in a different context. Moreover, the change in the helicity states during the neutrino flavor oscillation is consistent with similar results and observations in the literature \cite{dvornikov2013neutrino, dvornikov2019neutrino, joshi2020neutrino, guzzo2005random, chauhan2005low}, however, the mathematical approach shown here is different. Even though the results in the present article are shown through some formal expressions, in principle those complicated expressions can be evaluated using numerical techniques. It is important here to emphasize that the change in the helicity states during the neutrino oscillation could in principle be used as a detection technique in GW. Moreover, it can also be used to probe the nature of the interaction between the neutrinos and various classical sources during the neutrino emissions in various astrophysical sources like supernovae. However, the probability of the helicity changing during neutrino oscillation completely depends on the coupling strength of the interaction between neutrinos and background classical sources. As a result, we expect that the helicity change during the neutrino oscillation due to a passage of gravitational wave at asymptotic infinity would be very small due to the very small value of $\kappa = \sqrt{8\pi G}$ and the strength of the metric perturbation. On the other hand, if we consider the interaction between GW and neutrinos near the supernova explosions, the strength of this coupling would be high which makes the probability of helicity changing in the neutrino oscillation during this period relatively higher.  

Although the present analysis depends on the Dirac nature of neutrinos, a similar analysis can also be applicable to the Majorana nature of neutrinos where one might see new emerging features like CP violation which is recently shown in \cite{Dixit:2022izn}. This analysis is beyond the scope of the present article as it requires a separate detailed computation altogether.

\section{Acknowledgement}
SM is supported by SERB-Core Research Grant (Project RD/0122-SERB000-044).

\bibliographystyle{unsrt}
\bibliography{draft}

\end{document}